\documentclass{article}
\usepackage[T1]{fontenc} 
\usepackage[utf8]{inputenc} 
\usepackage[lbd]{ismir}
\usepackage{amsmath,cite,url}
\usepackage{graphicx}
\usepackage{color}
\usepackage{amssymb} 
\usepackage{subcaption}

\newcommand{\yes}{\checkmark}
\newcommand{\partly}{(\checkmark)}
\newcommand{\no}{--}

\title{PANEL: An Open-Source, Self-Hosted Web Platform for Human Evaluation of Generative Models}

\multauthor
{Matteo Spanio$^1$ \hspace{1cm} Andrea Poltronieri$^2$ \hspace{1cm} Mart\'{\i}n Rocamora$^2$} {
  $^1$ University of Padova, Padova, Italy\\
  $^2$ Music Technology Group, Universitat Pompeu Fabra, Barcelona, Spain\\
  {\tt\small spanio@dei.unipd.it, andrea.poltronieri@upf.edu}
}

\def\authorname{M. Spanio, A. Poltronieri, and M. Rocamora}

\begin{document}

\maketitle
\begin{abstract}
Human judgement is the reference measure for evaluating generative models, yet the software used to collect it lags behing the methodology. 
Researchers adapt listening-test frameworks designed for perceptual protocols such as MUSHRA, rely on closed commercial survey platforms, or implement single-use web applications. 
Live arenas such as Chatbot Arena and Music Arena rank publicly deployed systems at scale, but do not support controlled comparisons of a laboratory's own models with its own participants. We present PANEL, an open-source, self-hosted platform for such studies. 
A study is authored in the browser and distributed as a single link, with audio, video, image, and text stimuli, seven question types, and screening and skip logic. 
The platform reports per-question summaries, across-condition significance tests, pairwise win rates and Bradley--Terry scores, and supports power analysis from pilot data. 
Consent versioning, self-service withdrawal, retention enforcement, and audit logging support GDPR-compliant operation. Each study exports as a machine-readable specification.
PANEL is available at \url{https://github.com/matteospanio/panel}.
\end{abstract}
\section{Introduction}\label{sec:introduction}

\begin{figure}[ht]
 \centering
 \begin{subfigure}[t]{0.49\columnwidth}
  \centering
  \includegraphics[width=\linewidth]{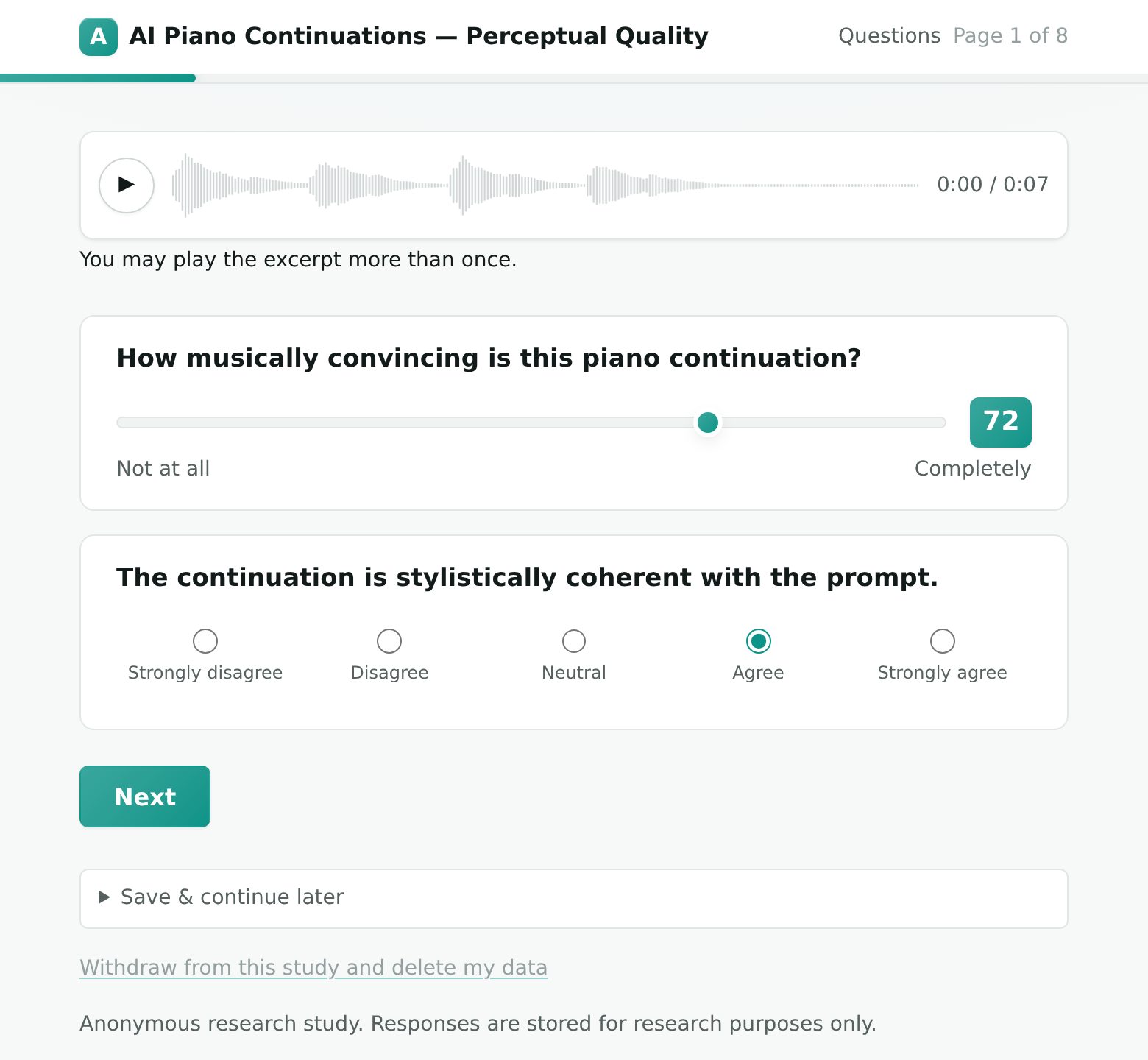}
  \caption{Participant view.}
  \label{fig:panel-participant}
 \end{subfigure}
 \hfill
 \begin{subfigure}[t]{0.49\columnwidth}
  \centering
  \includegraphics[width=\linewidth]{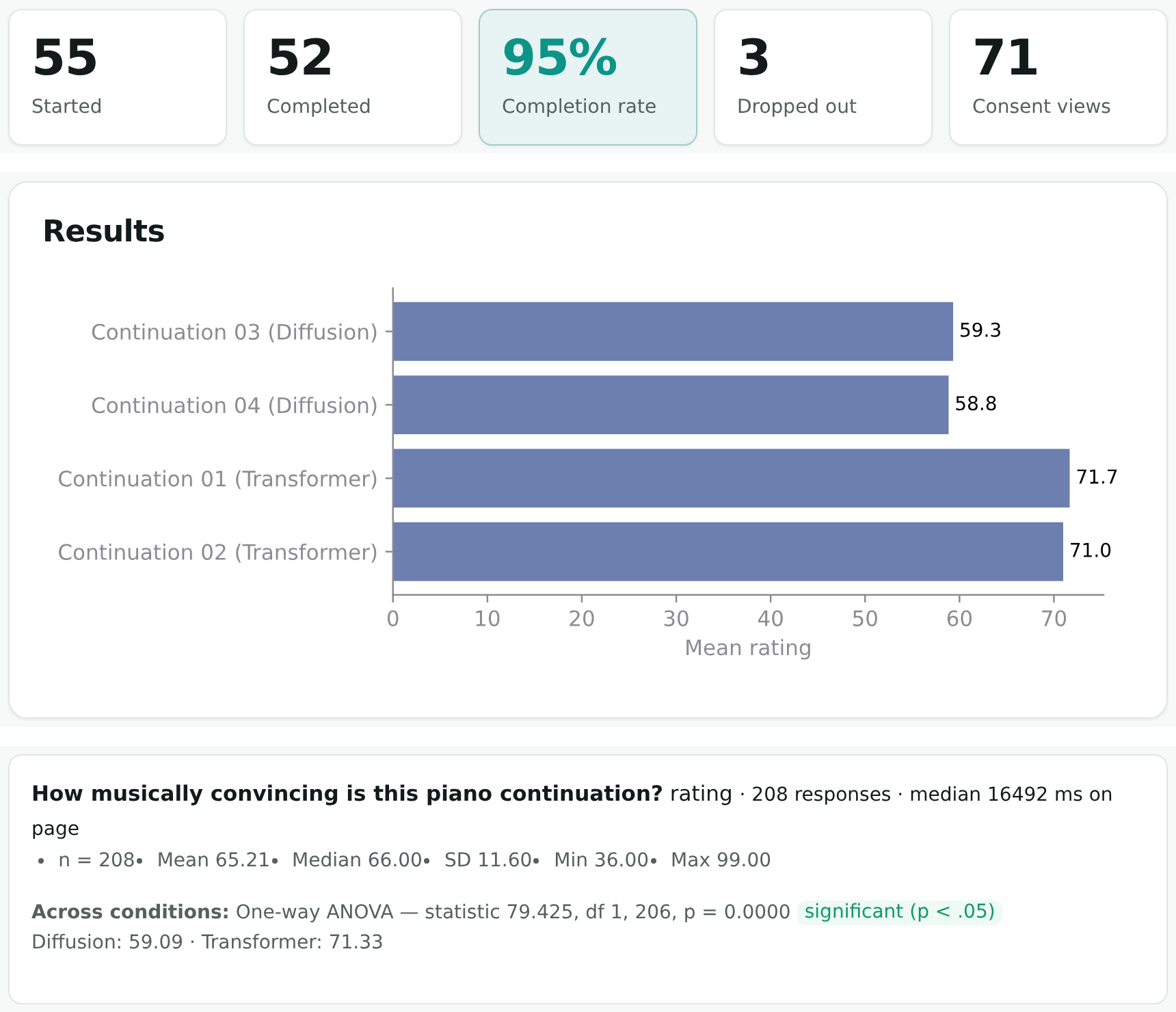}
  \caption{Researcher studio.}
  \label{fig:panel-studio}
 \end{subfigure}
 \caption{A demo study comparing two music-generation systems: (a) the participant view, with persistent save-and-resume and data-withdrawal links, and (b) the researcher studio (synthetic pilot data), with participation funnel, per-stimulus mean ratings, and an across-condition test.}
 \label{fig:panel}
\end{figure}

Generative models are ultimately evaluated by people. 
The objective metrics available to us correlate only weakly with what listeners report \cite{vinay2022evaluating}, and no automatic measure has displaced human listening as the arbiter of whether a system sounds good. In practice, text-to-music systems are therefore validated with listening studies \cite{agostinelli2023musiclm,copet2023simple}; surveys of the field treat subjective evaluation as the primary methodology \cite{yang2020evaluation,ji2024survey}.
The pattern extends beyond music: crowd-sourced pairwise preference is the standard against which automatic judges of language models are measured \cite{zheng2023judging}.

Collecting such judgments is as much an experimental-design problem as an audio one: per-system conditions, balanced stimulus assignment, screening and demographics, attention checks, inferential statistics, and documented handling of personal data. The audio community's listening-test frameworks (webMUSHRA \cite{schoeffler2018webmushra}, BeaqleJS \cite{kraft2014beaqlejs}, Go~Listen \cite{barry2021golisten}) implement ITU-R BS.1534 and related protocols \cite{itu2015bs1534} well, but are organised around the perceptual comparison rather than the surrounding experiment. Qualtrics \cite{qualtrics} and PsyToolkit \cite{stoet2017psytoolkit} provide the survey machinery, but neither can be deployed by the researcher: Qualtrics is a closed commercial service, and PsyToolkit's online interface is licensed non-commercially and runs from a single institutional server. 
In both cases responses reside on infrastructure the researcher does not control, so where personal data is stored and when it is erased are not theirs to determine.
Platforms built for model evaluation have arrived as live arenas: Chatbot Arena for language models \cite{chiang2024chatbot}, Music Arena for text-to-music \cite{kim2025music}. 
Each is one central service ranking public systems, which leaves no route for a laboratory to compare its own models, on its own stimuli, with participants it recruits itself.

We present \textbf{PANEL} (Platform for ANnotation \& EvaLuation), an open-source evaluation platform that a laboratory installs on its own server.
The platform, its deployment guides, and the study specification format are released under an open MIT licence.
Studies are authored in the browser, distributed as a single link, and analysed continuously as anonymous participants respond on their own devices (Figure~\ref{fig:panel}).
The platform has been used to run the subjective evaluation reported in \cite{poltronieri2026notation}; we describe its design here and invite the community to use and extend it.

\section{Related Tools}\label{sec:related}

Table~\ref{tab:tools} compares the tools named above along the requirements of controlled model-evaluation studies. PsychoPy \cite{peirce2019psychopy2} is the closest open-source relative in scope: an experiment generator whose Builder produces a program presenting stimuli with sub-millisecond timing precision in the laboratory \cite{bridges2020timing}. Reaching a browser requires compiling it to JavaScript and hosting it, in practice on Pavlovia, a paid service run by the maintainers (\pounds0.24 per participant) \cite{pavlovia2026}, which then holds recruitment, results, and governance. PANEL makes the opposite trade-off: a multi-user web service on infrastructure the laboratory controls, at the cost of precise stimulus-onset timing.
The studies we target do not need it -- a listener rating a multi-second excerpt is not making a reaction-time judgement -- and the browser cannot guarantee it in any case.

Live arenas invert the model again: anonymous visitors compare publicly deployed systems and a leaderboard aggregates the votes \cite{chiang2024chatbot,kim2025music}. 
This answers which released system is preferred overall; it does not support comparing two ablations among screened participants under a design with known statistical power.

\begin{table}
 \begin{center}
 \footnotesize
 \setlength{\tabcolsep}{2.7pt}
 \begin{tabular}{lccccc}
  \hline
   & Open & Self- & Beyond & Survey & Built-in \\
   & source & hosted & audio & logic & stats \\
  \hline
  webMUSHRA \cite{schoeffler2018webmushra} & \partly & \yes & \no & \partly & \no \\
  BeaqleJS \cite{kraft2014beaqlejs} & \yes & \yes & \no & \no & \no \\
  Go Listen \cite{barry2021golisten} & \yes & \yes & \no & \partly & \no \\
  PsyToolkit \cite{stoet2017psytoolkit} & \no & \no & \yes & \yes & \partly \\
  Qualtrics \cite{qualtrics} & \no & \no & \yes & \yes & \yes \\
  PsychoPy \cite{peirce2019psychopy2} & \yes & \partly & \yes & \yes & \no \\
  Arenas \cite{chiang2024chatbot,kim2025music} & \yes & \partly & \partly & \no & \yes \\
  PANEL & \yes & \yes & \yes & \yes & \yes \\
  \hline
 \end{tabular}
\end{center}
 \caption{\yes\ = yes, \partly\ = partial, \no\ = no. \emph{Survey logic}: screening, demographics, answer-dependent branching. \emph{Built-in stats}: significance tests or rankings computed in the tool. webMUSHRA (source-available, custom licence) and Go Listen have questionnaire pages but no branching; PsychoPy reaches the browser through PsychoJS, normally hosted on Pavlovia \cite{pavlovia2026}; arenas cover a single modality.}
 \label{tab:tools}
\end{table}

\section{Study Design}\label{sec:design}

A study consists of \emph{conditions} (typically one per model, checkpoint, or ablation), each holding \emph{stimuli}, and \emph{questions} participants answer about them. Besides audio, stimuli may be video, images, text, researcher-written HTML, or embedded URLs.

Two presentation modes are provided. In \emph{single-stimulus} mode each participant rates one item at a time; an assignment strategy determines how many items each person sees and in what order (balanced random, blocked, or counterbalanced); alternatively, a between-subject design assigns each participant a single condition. 
In \emph{pairwise} mode two outputs of the same prompt appear side by side, summarised as win rates and Bradley--Terry scores.

Seven question types (rating, choice, free text, Likert, numeric, matrix, ranking) are grouped into pages by author-placed page breaks \cite{stoet2017psytoolkit}, and run per stimulus, as demographics, or as \emph{screening}, whose eligibility rule terminates the session for ineligible participants. Questions may depend on earlier answers; these rules are evaluated server-side, so participants cannot inspect or bypass the branching and screening logic.

Quality controls follow standard survey practice. Questions can carry attention checks; sessions are flagged for implausibly fast completion, straight-lined scales, or duplicate participation; and per-stimulus playback time identifies ratings given without listening to the full excerpt. 
Flagged sessions can be excluded from all exports. Studies move through draft, pilot, active, and closed states: readiness checks gate activation, the structure is frozen once collection starts, and pilot responses are kept apart from the final dataset.
Studies can also be chained into timed phases for longitudinal designs.

\section{Analysis and Export}\label{sec:results}

Results are computed continuously over submitted, non-pilot sessions: distributions for choice and Likert items, summary statistics for ratings, mean ranks for rankings, win rates and Bradley--Terry scores for pairwise comparisons, each with the median response time. For per-stimulus questions an across-condition comparison is reported by default, with a test appropriate to the assignment strategy; in designs where participants rate multiple stimuli, the repeated measurements are accounted for rather than treated as independent. Any result can be segmented by device, country, condition, or week, with segmented comparisons marked as exploratory. A power calculator estimates required sample sizes from an expected effect size, or from pilot data as a first approximation.

Answers, demographics, pairwise comparisons, and the event log export as CSV; a scoped REST API and signed webhooks integrate with external pipelines. Each study also exports as a reproducibility bundle (conditions, stimuli with checksums, questions, and consent text), so the instrument can be inspected alongside its results.

\section{Extensibility and Deployment}\label{sec:implementation}

New question types and assignment strategies are Python classes registered with a decorator, discovered at startup, and shown in the study builder alongside the built-ins. Studies can be cloned and questions stored in a reusable bank. The platform deploys as a single Django service: SQLite for piloting, a bundled Docker Compose stack with PostgreSQL for production.

Self-hosting makes the laboratory the data controller for the responses it collects, and the platform supports the corresponding obligations: IP addresses are resolved to a country code offline and never stored; sessions record a hash of the consent text accepted; exports redact free-text fields marked as personal; retention windows are enforced per study; participants can erase their data at any time; and an append-only log records every edit and export.



\section{AI Usage Statement}

This manuscript was drafted with the assistance of a large language model (Claude, Anthropic), which was used for literature search, prose drafting, and the preparation of the demo study shown in the figures. All system descriptions, positioning claims, and references were verified by the authors, who take full responsibility for the content. The platform itself was developed with AI-assisted programming tools.

\bibliography{references}

\end{document}